# How (not) to assess the importance of correlations for the matching of spontaneous and evoked activity: a response

Michael Okun, Pierre Yger and Kenneth D. Harris

Institute of Neurology, Department of Neuroscience, Physiology, and Pharmacology, 21 University Street, University College London, London WC1E 6DE, UK

## Introduction

Neuronal population activity is often characterized experimentally by the distribution of multineuron "words" (binary firing vectors). Previous studies have described ways to analyse the structure of multineuron word distributions in terms of pairwise correlations between individual spike trains. In recent work, we approached this question from an alternative perspective, in which population rate, rather than pairwise correlations is used to explain the word distribution structure [1]. We found that the structure of word distributions in sensory cortex is dominated by population rate fluctuations. We also used this approach to re-evaluate the case for sampling-based representation and inference in primary sensory cortex, made in [2]. A response to our publication by the authors of [2] has recently become available [3]. Although the new analysis and several other issues addressed by [3] are of considerable interest, other parts of their response are based on a misinterpretation of our results and claims. We maintain that our original analysis was correct, from both mathematical and biological perspectives. Here we address the points raised in [3] and summarize, to the best of our understanding, which issues both groups agree on, and which questions remain disputed.

## Information-theoretic considerations

Central to the entire argument is the concept of synthetic distributions. Consistently with the notation used in [2,1], for an empirically observed word distribution $P$, we use $\tilde{P}$ to denote the synthetic distribution in which each spike train retains its rate, but is independent of all other spike trains. In [1] we introduced the more elaborate raster marginals model (RMM) for word distributions, which retains both the firing rates and the population rate distribution of the original, but otherwise is as random as possible (see Fig. 2B in [1]). As in [1], we use $\hat{P}$ to denote the RMM word distribution corresponding to $P$ (in [3] this was denoted by $\tilde{P}_{Okun}$). The similarity of two word distributions $P$ and $Q$ is quantified by Kullback-Leibler divergence (KLdiv) between them, denoted by $D[P||Q]$. This quantity is estimated as described in [2,1].

In the following we consider the specific points raised in [3], in the order they appear.

**1.** The match between $P$ and $\hat{P}$.

The point of [1] was *not* that $P$ and $\hat{P}$ are identical. Indeed, right after introducing RMM, we explained that it does not capture specific correlations that might be present in the data (see Fig. 2C in [1]). We next proceeded to show how in electrophysiological data from rat A1 the actual pairwise correlations differ from those predicted by RMM (Fig. 3C in [1]). This fact was very clearly stated in [1] (e.g. "although the RMM greatly outperforms prediction from mean firing rates, it still does not provide a perfect fit to the observed word distributions".) In other words, we do not claim that all correlations between neurons come from fluctuations in population rate. Rather, we claim that because population rate fluctuations can exert such a large effect on correlations, they must be understood and modelled before any claims are made regarding learning of environmental models.

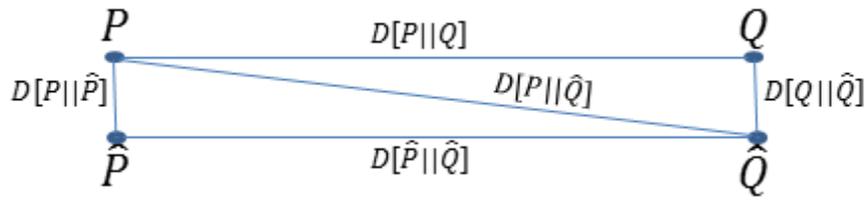

**Figure 1.** Illustration of the relative similarities of original and RMM-scrambled data.

**2.** Using $D[\widehat{P}||\widehat{Q}]$ as a proxy for $D[P||Q]$.

In [3] it was observed that if $P = Q$ then, by mathematical necessity, $D[\widehat{P}||\widehat{Q}] = 0$. This is of course correct, but is unrelated to the point made in [1].

In [1], we made an empirical observation that $D[\widehat{P}||\widehat{Q}]$ and $D[P||Q]$ closely match[1], not only for pairs of distributions $P$ and $Q$ for which $D[P||Q]$ is low, but also for pairs with high $D[P||Q]$. This is not a mathematical necessity, but rather a property specific to word distributions recorded in the sensory cortex. Furthermore, the match of $D[\widehat{P}||\widehat{Q}]$ and $D[P||Q]$ relies on the synthetic rasters being produced by RMM, and it does not necessarily hold for other ways of synthesizing surrogate rasters. For instance, it was shown in [1] that $D[\widetilde{P}||\widetilde{Q}]$ and $D[P||Q]$ do not match (Figs. 4D, 5D). Importantly, these empirical observations, which we made with data recorded in rat A1 and cat V1, were confirmed in [3] for ferret V1 data. Specifically, Fig. B of [3] shows that $D[\widehat{P}||\widehat{Q}]$ and $D[P||Q]$ closely match across all developmental stages (red vs purple bars), whereas $D[\widetilde{P}||\widetilde{Q}]$ and $D[P||Q]$ match only in adults (red vs pink bars).

The relationship of $D[P||\widehat{Q}]$ and $D[P||Q]$ was not central to the analysis of [1], and was therefore discussed only briefly. However, contrary to the suggestion in [3], we did not claim that the two divergences match. Indeed, following Fig. 4C of [1], which shows a graph of $D[P||\widehat{Q}]$ vs. $D[P||Q]$ in rat A1 data, we noted that $D[P||\widehat{Q}]$ is slightly but consistently larger than $D[P||Q]$, "with a small additive offset approximately equal to the residual distance between the original and model data".

These observations are summarized in an intuitive but informal way in Fig. 1. In our data, the divergence $D[P||Q]$ between word distributions corresponding to two experimental conditions (e.g. spontaneous and evoked activity) could be large or small, depending on the particular stimulus and population recorded. In all cases, however, $D[\widehat{P}||\widehat{Q}]$ was very similar to $D[P||Q]$. Consistent with the intuition of Fig. 1, $D[P||\widehat{Q}]$ was greater than $D[P||Q]$. These observations suggest an informal conjecture, that the differences between $P$ and $Q$ are in some sense orthogonal to the differences between $P$ and $\hat{P}$. Formalization of this conjecture might be possible for example using methods of Information Geometry [4].

---

[1] We however do *not* mean that the match between $D[\widehat{P}||\widehat{Q}]$ and $D[P||Q]$ is absolutely precise, in the sense that no statistically significant difference between the two exists.

## Biological interpretation

Since publication of [1], Fiser et al. have applied the RMM to ferret V1 data, reporting the results in [3]. This analysis – which shows a statistically significant difference between $D[M||S]$ and $D[M||\hat{S}]$ (where $M, S$ denote word distributions during natural movie presentation and spontaneous conditions) – indeed suggests that the similarity of $M$ and $S$ is not simply a consequence of similar population rate dynamics. Nevertheless, this still does not indicate that an environmental model has been learned. To see why, consider the random recurrent network shown in Fig. 8 of [1], in condition 2, where input driven and spontaneous activity is similar. In this condition $D[Id||S]$=45 bits/s, whereas $D[Id||\hat{S}]$=74 bits/s (where $Id, S$ denote word distributions during input driven and spontaneous conditions; this result was not included in [1]). Thus, the match between $Id$ and $S$ is due to both similarity of population dynamics, and to the match of specific correlations between neurons. In this randomly-connected network, however, this is not a signature of learning, but merely of the network connectivity being the same in $Id$ and $S$ conditions.

In [1], we suggested that the major changes in population word distributions observed across development might arise not from learning, but from developmental changes in network dynamics. This was supported by the fact that we were able to produce such behaviour by simple adjustment in the parameters of a random spiking network model, which exhibited no synaptic plasticity or other learning. In an interesting extension of their original study, Fiser et al. recently presented data comparing the match between spontaneous and evoked activity in ferrets of different ages whose eyelids were kept sutured until the very beginning of the experiment [5,6]. They reported that in these animals, evoked and spontaneous activity also become significantly more similar with development, although some dissimilarity remains even in adults. This direct experimental evidence demonstrates that developmental factors not related to learning are a major factor in shaping the structure of cortical activity.

Characterizing the structure of multineuron spike patterns is a very important challenge for systems neuroscience. Developing quantitative tools for analysis of word distributions is an important step toward this goal, but it is equally important to understand the caveats of any particular method, to ensure that accurate biological conclusions are drawn. If it were shown that a sampling-based representation is indeed employed in cortex, this would constitute a major insight into cortical function. While we maintain that [2] did not provide conclusive evidence for this hypothesis, it served an extremely valuable contribution of clearly formulating this hypothesis and bringing it to the attention of a wide audience. Although the debate is not yet settled, more recent results [1, 3, 5, 6] have provided new insights into the question.